\documentclass[final]{IEEEtran}
\IEEEoverridecommandlockouts
\usepackage{cite}
\usepackage{amsmath,amssymb,amsfonts}
\usepackage{algorithmic}
\usepackage{graphicx}
\usepackage{textcomp}
\usepackage{xcolor}
\usepackage{array}

\def\BibTeX{{\rm B\kern-.05em{\sc i\kern-.025em b}\kern-.08em
    T\kern-.1667em\lower.7ex\hbox{E}\kern-.125emX}}
\begin{document}

\title{Survey of Machine Learning Accelerators\\
\thanks{This material is based upon work supported by the Assistant Secretary of Defense for Research and Engineering under Air Force Contract No. FA8721-05-C-0002 and/or FA8702-15-D-0001. Any opinions, findings, conclusions or recommendations expressed in this material are those of the author(s) and do not necessarily reflect the views of the Assistant Secretary of Defense for Research and Engineering.}
}

\author{\IEEEauthorblockN{Albert Reuther, Peter Michaleas, Michael Jones, Vijay Gadepally, Siddharth Samsi, and Jeremy Kepner} \\
\IEEEauthorblockA{\textit{MIT Lincoln Laboratory Supercomputing Center} \\
Lexington, MA, USA \\
\{reuther,pmichaleas,michael.jones,vijayg,sid,kepner\}@ll.mit.edu}
}

\maketitle

\begin{abstract}

New machine learning accelerators are being announced and released each month for a variety of applications from speech recognition, video object detection, assisted driving, and many data center applications. This paper updates the survey of of AI accelerators and processors from last year's IEEE-HPEC paper. This paper collects and summarizes the current accelerators that have been publicly announced with performance and power consumption numbers. The performance and power values are plotted on a scatter graph and a number of dimensions and observations from the trends on this plot are discussed and analyzed. For instance, there are interesting trends in the plot regarding power consumption, numerical precision, and inference versus training. This year, there are many more announced accelerators that are implemented with many more architectures and technologies from vector engines, dataflow engines,  neuromorphic designs, flash-based analog memory processing, and photonic-based processing. 

\end{abstract}


\section{Introduction}

It has become apparent that researching, developing and deploying Artificial Intelligence (AI) and machine learning (ML) solutions has become a promising path to addressing the challenges of evolving events, data deluge, and rapid courses of action faced by many industries, militaries, and other organizations. Advances in the past decade in computations, data sets, and algorithms have driven many advances for machine learning and its application to many different areas. 

\begin{figure}[th]
    \centering
    \includegraphics[width=3in]{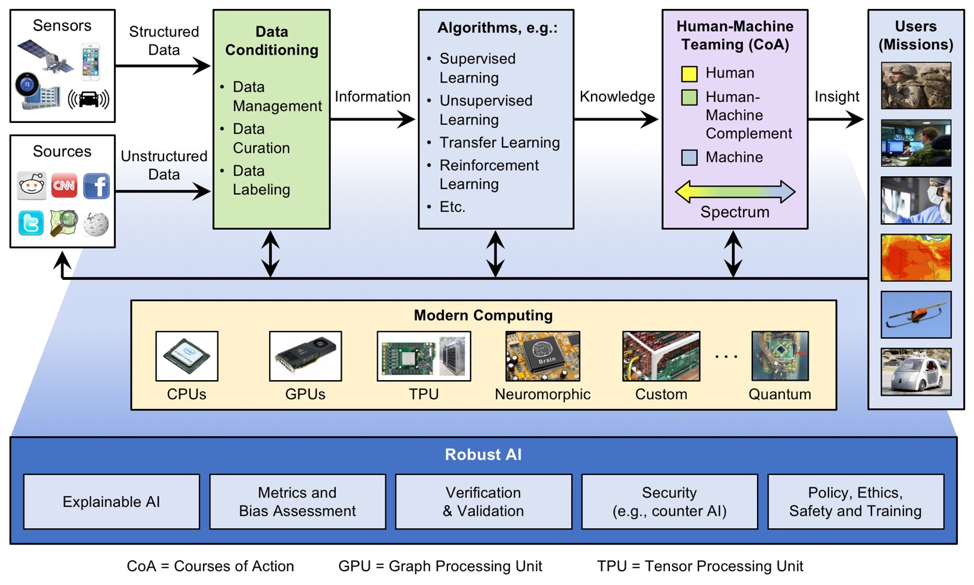}
    \caption{Canonical AI architecture consists of sensors, data conditioning, algorithms, modern computing, robust AI, human-machine teaming, and users (missions). Each step is critical in developing end-to-end AI applications and systems.}
    \label{fig:architecture}
  \end{figure}

AI systems bring together a number of components that must work together to effectively provide capabilities for use by decision makers, warfighters, and analysts~\cite{gadepally2019enabling}. 
Figure~\ref{fig:architecture} captures this system and its components of an end-to-end AI solution. 
On the left side of Figure~\ref{fig:architecture}, structured and unstructured data sources provide different views of entities and/or phenomenology. 
These raw data products are fed into a data conditioning step in which they are fused, aggregated, structured, accumulated, and converted to information. The information generated by the data conditioning step feeds into a host of supervised and unsupervised algorithms such as neural networks, which extract patterns, predict new events, fill in missing data, or look for similarities across datasets, thereby converting the input information to actionable knowledge. This actionable knowledge is then passed to human beings for decision-making processes in the human-machine teaming phase. The phase of human-machine teaming provides the users with useful and relevant insight turning knowledge into actionable intelligence or insight. 

Underpinning this system are modern computing systems, for which Moore's law trends have ended~\cite{theis2017end}, as have a number of related laws and trends including Denard's scaling (power density), clock frequency, core counts, instructions per clock cycle, and instructions per Joule (Koomey's law)~\cite{horowitz2014computing}. However, advancements and innovations are still progressing in the form of specialized circuits and chips that accelerate often-used operational kernels, methods, or functions.  These accelerators are designed with a different balance between performance and functional flexibility. This includes an explosion of innovation in ML processors and accelerators~\cite{hennessy2019new,dally2020domain}. Understanding the relative benefits of these technologies is of particular importance to applying AI to domains under significant constraints such as size, weight, and power, both in embedded applications and in data centers. 


This paper is an update to last year's IEEE-HPEC paper~\cite{reuther2019survey}. Quite a number more accelerator chips have been announced and released, and other technologies like neuromorphic architectures, memory-based analog acceleration, and computing with light are gaining attention. There are also some technology categories that were included in last year's paper that will not be included this year, namely most FPGA-based inference instances~\cite{li2017survey,mittal2018survey,blaiech2019survey,guo2019survey} and smartphone accelerators. Only FPGA-based offerings that have some dynamic programmability are considered in this paper (e.g., Intel Arria, AImotive, and Microsoft Brainwave). Smartphone accelerators are not being considered this year because they cannot be used in a different platform without significant re-engineering. 

Before getting to the accelerators, we will review a few topics pertinent to understanding the capabilities of the accelerators. We must discuss the types of neural networks for which these ML accelerators are being designed; the distinction between neural network training and inference; the numerical precision with which the neural networks are being used for training and inference, and how neuromorphic accelerators fit into the mix: 

\begin{itemize}

\item Types of Neural Networks -- While AI and machine learning encompass a wide set of statistics-based technologies~\cite{gadepally2019enabling}, this paper continues with last year's focus on processors that are geared toward deep neural networks (DNNs) and convolutional neural networks (CNNs).  Overall, the most emphasis of computational capability for machine learning is on DNN and CNNs because they are quite computationally intensive~\cite{canziani2016analysis}, and because most of the computations are dense matrix-matrix and matrix-vector multiplies, they are primed to take advantage of computational architectures that exploit data reuse, data locality, and data density. 

\item Neural Network Training versus Inference -- As was explained in last year's survey, neural network training uses libraries of input data to converge model weight parameters by applying the labeled input data (forward projection), measuring the output predictions and then adjusting the model weight parameters to better predict output predictions (back projection).  Neural network inference is using a trained model of weight parameters and applying it to input data to receive output predictions. Processors designed for training may also perform well at inference, but the converse is not always true. 

\item Numerical Precision -- The numerical precision with which the model weight parameters and model input data are stored and computed has an impact on the accuracy and efficiency with which networks are trained and used for inference. Generally higher numerical precision representations, particularly floating point representations, are used for training, while lower numerical precision representations, (in particular, integer representations) have been shown to be reasonably effective for inference~\cite{sze2017efficient, narang2018mixed}. However, it is still an open research question whether very limited numerical precisions like int4, int2, and int1 adequately represent model weight parameters and significantly affect model output predictions. Over the past year, more training accelerators have been released that support 16-bit floating point numbers (fp16 and bfloat16), and most inference accelerators now release performance results for 8-bit integer (int8) operands, particularly the inference accelerators geared at edge and embedded processing applications.  
\item Neuromorphic Computing -- The field of neuromophic computing emerged from the neuroscience field, in which researchers and engineers design circuits to model biological and physiological mechanisms in brains. Schuman's survey~\cite{schuman2017survey} provides a rich background of all of the significant efforts in the field over the past several decades. Some of the most prominent features of neuromorphic circuits are synthetic neurons and synapses along with spiking synapse signaling, and many agree that the spiking synapse signal is the most prominent feature, especially when implemented in neural network accelerators. 
In recent accelerators, these spiking signals are usually encoded as digital signals, e.g., IBM TrueNorth chip~\cite{merolla2014million,esser2016convolutional}, University of Manchester SpiNNaker~\cite{khan2008spinnaker}, and Intel's Loihi~\cite{lin2018programming}. Another recent notable research accelerator is the Tianjic research chip~\cite{pei2019towards}, developed by a team at Tsinghua University, which demonstrated the capability of choosing either a spiking neural network (SNN) layer or non-spiking artificial neural network (ANN) layer to instantiate each layer of a DNN model for inference. The team showed that for certain models for audio and video processing a hybrid layer-by-layer approach was most effective, both in accuracy and power consumption. Finally, a startup called Knowm is developing a new neuromorphic computational framework called AHaH Computing (Anti-Hebbian and Hebbian) based on memristor technology~\cite{nugent2014ahah}. Their goal is to use this technology to dramatically reduce SWaP for machine learning applications. 
\end{itemize}

There are many surveys~\cite{lindsey1995survey,liao2001neural,misra2010artificial,sze2017efficient,sze2020efficient,langroudi2019digital,chen2020survey,wang2019deep,khan2020ai,rueckert2020digital} and other papers that cover various aspects of AI accelerators; this paper focuses on gathering a comprehensive list of AI accelerators with their computational capability, power efficiency, and ultimately the computational effectiveness of utilizing accelerators in embedded and data center applications, as did last year's paper. Along with this focus, this paper mainly compares neural network accelerators that are useful for government and industrial sensor and data processing applications. Therefore, it will make a distinction between research accelerators and commercially available accelerators, and it will focus comparisons on the latter. Research accelerators are developed by university and industry researchers to demonstrate the art of the possible, often with the intent of influencing commercial attention to technology developments or to attract venture funding to commercialize their developed technology. But before either of these intents are realized, the demonstration of the art of the possible show us opportunities in which commercial products may pursue for improved capabilities and features.

\section{Survey of Processors}

\begin{figure*}[htb]
    \includegraphics[width=\textwidth]{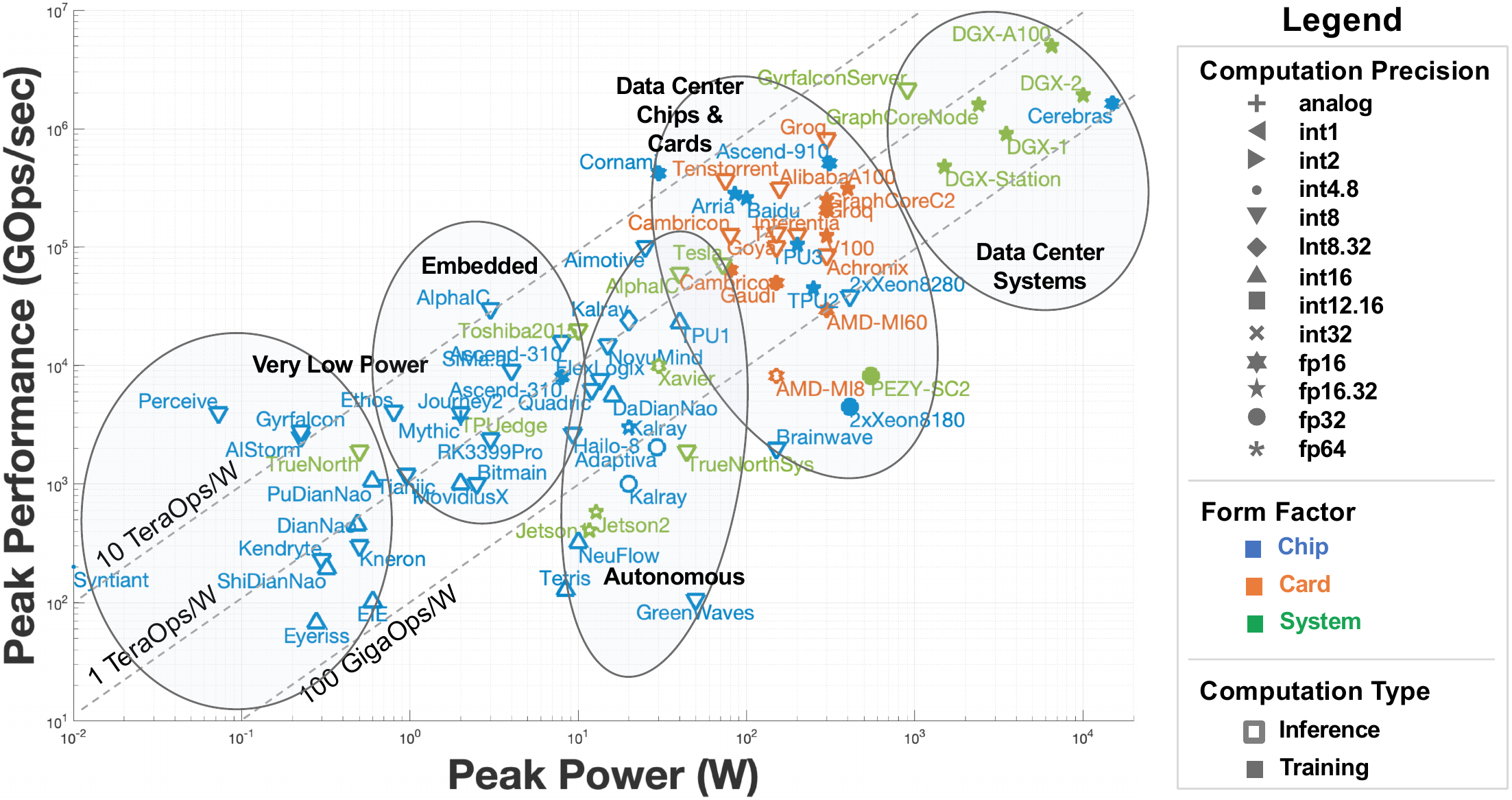}
    \caption{Peak performance vs. power scatter plot of publicly announced AI accelerators and processors.}
    \label{fig:PeakPerformancePower}
  \end{figure*}

Many recent advances in AI can be at least partly credited to advances in computing hardware~\cite{krizhevsky2012imagenet,jouppi2018domain}, enabling computationally heavy machine-learning algorithms such as neural networks. This survey gathers performance and power information from publicly available materials including research papers, technical trade press, company benchmarks, etc. While there are ways to access information from companies and startups (including those in their silent period), this information is intentionally left out of this survey; such data will be included in this survey when it becomes publicly available. The key metrics of this public data are plotted in Figure~\ref{fig:PeakPerformancePower}, which graphs recent processor capabilities (as of June 2020) mapping peak performance vs. power consumption. 
The x-axis indicates peak power, and the y-axis indicate peak giga-operations per second (GOps/s). Note the legend on the right, which indicates various parameters used to differentiate computing techniques and technologies. The computational precision of the processing capability is depicted by the geometric shape used; the computational precision spans from analog and single-bit int1 to four-byte int32 and two-byte fp16 to eight-byte fp64. The precisions that show two types denotes the precision of the multiplication operations on the left and the precision of the accumulate/addition operations on the right (for example, fp16.32 corresponds to fp16 for multiplication and fp32 for accumulate/add). The form factor is depicted by  color; this is important for showing how much power is consumed, but also how much computation can be packed onto a single chip, a single PCI card, and a full system. Blue corresponds to a single chip; orange corresponds to a card (note that they all are in the 200-300 Watt zone); and green corresponds to entire systems (single node desktop and server systems). This survey is limited to single motherboard, single memory-space systems. Finally, the hollow geometric objects are peak performance for inference-only accelerators, while the solid geometric figures are performance for accelerators that are designed to perform both training and inference. 

We can make some general observations from Figure~\ref{fig:PeakPerformancePower}. First, quite a number of new accelerator chips, cards, and systems have been announced and released in the past year. Each of the five categories have a higher density of entries from last year, and there is a greater diversity of underlying architectures and technologies with which these accelerators have been developed as you will notice in the descriptions below. Also, many more recent accelerators have broken through the 1 TeraOps/W boundary for peak performance. 
An observation that has not changed from last year is that at least 100W must be employed to perform training; almost all of the points on the scatter plot below 100W are inference-only processors/accelerators. (Cornami is the one exception, but their plot point is based on simulation estimates.) It is generally understood that training requires floating point precision which requires more capable and power-consuming ALUs, datapaths, memories, etc.  

When it comes to precision, there is an even wider variety of numerical precisions for which peak performance numbers have been released. There continues to be exploration about how much precision is necessary for neural networks to perform well even when using limited or mixed precision representation of activation functions, weights, and biases~\cite{narang2018mixed,gupta2015deep}. 

Finally, a reasonable categorization of accelerators follows their intended application, and the five categories are shown as ellipses on the graph, which roughly correspond to performance and power consumption: Very Low Power for speech processing, very small sensors, etc.; Embedded for cameras, small UAVs and robots, etc.; Autonomous for driver assist services, autonomous driving, and autonomous robots; Data Center Chips and Cards; and Data Center Systems. In the following listings, the angle-bracketed string is the label of the item on the scatter plot, and the square bracket after the angle bracket is the literature reference from which the performance and power values came. A few of the performance values are reported in frames per second (fps) with a given machine learning model. For translating fps values to performance values, Samuel Albanie's Matlab code and a web site of all of the major machine learning models with their operations per epoch/inference, parameter memory, feature memory, and input size~\cite{albanie2019convnet} were used. 

\subsection{Research Chips}

Many research papers have been published that have introduced, evaluated, and compared various architectural elements, organizations, and technologies. The following list contains just a fraction of the research chips, but most of these have been highly cited. And, they have performance and power numbers. 

\begin{itemize}
\item The NeuFlow chip $\langle$NeuFlow$\rangle$~\cite{farabet2011neuflow} was a project between IBM, New York University and Yale University to explore architectures for efficient edge inference. It was designed with dataflow processing tiles comprised of a matrix multiplier and convolver in a 2-dimensional grid. 
\item The Stanford-designed energy efficient inference engine (EIE) $\langle$EIE$\rangle$~\cite{han2016eie} demonstrated the use of sparsity in DNN with compressed neural network models, weight sharing, and on-chip SRAM utilization to design an extremely efficent inference chip. 
\item Another Stanford chip, the TETRIS $\langle$TETRIS$\rangle$~\cite{gao2017tetris}, demonstrated highly efficient inference by using 3-dimensional memory and a partitioning scheme that placed weights and data close to their processing elements. This allowed the team to use more chip area for computation than local SRAM memory. 
\item MIT Eyeriss chip $\langle$Eyeriss$\rangle$~\cite{chen2018eyeriss,chen2017eyeriss,sze2017efficient} is a research chip from Vivienne Sze's group in MIT CSAIL. Their goal was to develop the most energy efficient inference chip possible by experimenting with different circuit computation trade-offs. The reported result was acquired running AlexNet with no mention of batch size. 
\item The DianNao series of dataflow research chips came from a university research team primarily at the Institute of Computing Technology of the Chinese Academy of Sciences (IST-CAS); this team overlaps with the Cambricon company, which has released several chips as well as the Kirin accelerator that is included in the Huawei smartphone system-on-chip (SoC). They published four different designs aimed at different types of ML processing~\cite{chen2016diannao}. The DianNao $\langle$DianNao$\rangle$~\cite{chen2016diannao} is a neural network inference accelerator, and the DaDianNao $\langle$DaDianNao$\rangle$~\cite{chen2014dadiannao} is a many-tile version of the DianNao for larger NN model inference. The ShiDianNao $\langle$ShiDianNao$\rangle$~\cite{du2015shidiannao} is designed specifically for convolutional neural network inference. Finally, the PuDianNao $\langle$PuDianNao$\rangle$~\cite{liu2015pudiannao} is designed for seven representative machine learning techniques: k-means, k-NN, na\"ive Bayes, support vector machines, linear regression, classification tree, and deep neural networks. 
\item The TrueNorth $\langle$TrueNorth$\rangle$~\cite{esser2016convolutional,akopyan2015truenorth,feldman2016ibm} is a digital neuromorphic research chip from the IBM Almaden research lab. It was developed under DARPA funding in the Synapse program to demonstrate the efficacy of digital spiking neural network (neuromorphic) chips. Note that there are points on the graph for both the system, which draws the 44 W power, and the chip, which itself only draws up to 275 mW. The TrueNorth has inspired several companies to develop and release neuromorphic chips. 
\end{itemize}

\subsection{Very Low Power Chips}

In the past year, many new chips have been announced or are offering inference-only products in this space.  

\begin{itemize}
\item The Intel MovidiusX processor $\langle$MovidiusX$\rangle$~\cite{hruska2017new} is an embedded video processor that includes a Neural Engine for video processing and object detection. 
\item In early 2019, Google released a TPU Edge processor $\langle$TPUEdge$\rangle$~\cite{tpu2019edge} for embedded inference application. The TPU Edge uses TensorFlow Lite, which encodes the neural network model with low precision parameters for inference.
\item The Rockchip RK3399Pro $\langle$Rockchip$\rangle$~\cite{rockchip2018rockchip} is an image and neural co-processor from Chinese company Rockchip. They published peak performance numbers for int8 inference for their ARM CPU and GPU-based co-processor. 
\item The GreenWaves GAP9 processor $\langle$GreenWaves$\rangle$~\cite{turley2020gap9,greenwaves2020gap} is designed for battery powered sensors and wearable devices including audio and video processing with respectable performance for such low power consumption. It is capable of computing in 8-, 16-, 24-, and 32-bit integer and 8-, 16-, and 32-bit floating point. 
\item The Kneron KL520 accelerator $\langle$Kneron$\rangle$~\cite{ward2020kneron} is designed for AI smart locks, smart home, people monitoring, and retail applications. The first released systems connect the KL520 with a dual-core ARM M4 processor.
\item San Jose startup AIStorm $\langle$AIStorm$\rangle$~\cite{merrit2019startup} claims to do some of the math of inference up at the sensor in the analog domain. They originally came to the embedded space scene with biometric sensors and processing, and they call their chip an AI-on-Sensor capability, using technology similar to Mythic and Syntiant, below. 
\item The Syntiant NDP101 chip $\langle$Syntiant$\rangle$~\cite{mcgrath2018tech,merrit2019startup} is the neural decision processor inside the Amazon Alexa products for attention phrase recognition. It is a processor-in-memory design, which performs inference weight computations in int4 and activation computation in int8 and consumes less than 200~{\textmu}W. 
\item The Mythic Intelligent Processing Unit accelerator $\langle$Mythic$\rangle$~\cite{fick2018mythic,hemsoth2018mythic} combines a RISC-V control processor, router, and flash memory that uses variable resistors (i.e., analog circuitry) to compute matrix multiplies. The accelerators are aiming for embedded, vision, and data center applications. 
\item The Perceive Ergo processor $\langle$Perceive$\rangle$~\cite{mcgregor2020perceive} is meant for battery-powered applications that require high performance. Ergo can run multiple concurrent networks with over 100 million weights in applications such as simultaneous video and audio processing. 
\item The XMOS Xcore.ai microcontroller $\langle$Xcore-ai$\rangle$~\cite{ward2020xmos} is a very low power processor that supports int1, int8, int16, and int32 computation. It is initially targeted for voice applications such as keyword detection and dictionary functions. 

\end{itemize}

\subsection{Embedded Chips and Systems} 

The systems in this category are aimed at embedded applications that require high performance inference relative to power and form factor including embedded camera processors, small UAVs, modest robots, etc.

\begin{itemize}
\item ARM has released its Ethos line of machine learning processors that mate one or more of its big.LITTLE ARM cores with MAC Compute Engines (MCEs). The Ethos N77 $\langle$Ethos$\rangle$~\cite{schor2020arm} has four MCEs, each of which are capable of computing 1 TOP/s at 1.0 GHz. 
\item Bitmain is a Chinese company that has specialized in cryptocurrency mining, but they are broadening their product lineup with the Bitmain BM1880 embedded AI processor $\langle$Bitmain$\rangle$~\cite{wheeler2019bitmain}. The processor features two ARM Cortex CPUs, a RISC-V CPU controller, and a tensor processing unit, which is aimed at video surveillance applications. 
\item The NovuMind NovuTensor chip $\langle$NovuMind$\rangle$~\cite{yoshida2018novumind,freund2019novumind} is a co-processor for vision inference applications. 
\item The Lightspeeur 5801 video processing chip from Gyrfalcon $\langle$Gyrfalcon$\rangle$~\cite{ward2019gyrfalcon} uses a matrix processing engine with processor-in-memory techniques to compute model inference. A server is available (see below) that incorporates 128 Lightspeeur 5801 chips as accelerators. 

\end{itemize}

\subsection{Autonomous Systems}

The entries in this category are aimed at inference processing for automotive AI/ML, autonomous vehicles, UAV/RPAs, robots, etc. 

\begin{itemize}
\item The AImotive aiWare3 $\langle$AImotive$\rangle$~\cite{aimotive2018aiware3} is a programmable FPGA-based accelerator aimed at the autonomous driving industry. 
\item The AlphaIC Real AI Processor-Edge (RAP-E) $\langle$AlphaIC$\rangle$~\cite{clarke2018indo} touts agent-based neural network computations, where agents are associated with each computational kernel that is executed.   
\item The Israeli startup Hailo released some performance details about its Hailo-8 edge inference chip in 2019 $\langle$Hailo-8$\rangle$~\cite{ward2019details}. They published ResNet-50 inference numbers but no peak performance number or architectural details from which peak numbers could be derived so the peak performance number on the graph comes from the ResNet-50 performance. 
\item Horizon Robotics has started producing its Journey2 accelerator $\langle$Journey2$\rangle$~\cite{horizon2020journey} for autonomous driving vision applications. 
\item The Huawei HiSilicon Ascend 310 $\langle$Ascend310$\rangle$~\cite{huawei2020ascend310} is an integrated CPU with AI accelerator based on the same Huawei Da Vinci architecture as the data center focused Ascend 910 (see below). While the Ascend 310 reports respectable performance for both int8 and fp16, it is intended for inference applications. 
\item The Kalray Coolidge $\langle$Kalray$\rangle$~\cite{dupont2019kalray,clarke2020nxp} parallel processor technology originally was developed for high performance computing for French nuclear simulations. It became apparent to Kalray that the same architecture could be used for AI inference tasks for automotive applications, which has led to a partnership with French chip manufacturer, NXP. The Coolidge chip has 80 64-bit VLIW procesor cores that each have an AI coprocessor. 
\item The NVIDIA Jetson-TX1 $\langle$JetsonTX1$\rangle$~\cite{franklin2017nvidia} incorporates four ARM cores and 256 CUDA Maxwell cores, and the Jetson-TX2 $\langle$JetsonTX2$\rangle$~\cite{franklin2017nvidia} mates six ARM cores with 256 CUDA Pascal cores. To gain more performance, the NVIDIA Xavier $\langle$Xavier$\rangle$~\cite{hruska2018nvidia} deploys eight ARM cores with 512 CUDA Volta cores and 64 Tensor cores. 
\item The Quadric q1-64 accelerator chip $\langle$Quadric$\rangle$~\cite{firu2019quadric} has a 4,096 fused multiply-add (FMA) array for inference processing. One Quadric card will contain four Quadric q1-64 chips, and they have plans for producing a q1-128 chip with a 16k FMA array for data center applications. 
\item The SiMa.ai Machine Learning Accelerator (MLA)  $\langle$Quadric$\rangle$~\cite{firu2019quadric} combines an ARM CPU with a dataflow matrix multiply engine to achieve high computation rates with only 4 Watts. 
\item The Tesla Full Self-Driving (FSD) Computer chip $\langle$Tesla$\rangle$~\cite{wikichip2020fsd,talpes2020compute} has two neural processing units (NPUs); each NPU has a 96x96 MAC array with 8-bit multiply and 32-bit add units. The FSD also has a set of 12 ARM Cortex-A72 CPUs and a GPU, but the bulk of the computational capability is delivered by the NPUs. 
\item The Toshiba 2015 image processor $\langle$Toshiba2015$\rangle$~\cite{merritt2019samsung} combines two 4-core ARM Cortex-A53 processors, four DSPs, a DNN accelerator, and several application specific accelerators, and it is designed for the autonomous driving market. 
\end{itemize}

\subsection{Data Center Chips and Cards}

There are a variety of technologies in this category including several CPUs, a number of GPUs, programmable FPGA solutions, and dataflow accelerators. They are addressed in their own subsections to group similar processing technologies. 

\subsubsection{CPU-based Processors}
\begin{itemize}
\item The Intel second-generation Xeon Scalable processors $\langle$2xXeon8180$\rangle$ and $\langle$2xXeon8280$\rangle$~\cite{degelas2019intel} are conventional high performance Xeon server processors. Intel has been marketing these chips to data analytics companies as very versatile inference engines with reasonable power budgets. The peak performance is computed as the peak of two Intel Xeon Platinum CPUs (8180 for inference and 8280 for training) with the key math units being the dual AVX-512 units per core. 
\item The Japanese Pezy SC2 massively multicore processor $\langle$Pezy$\rangle$~\cite{schor2017pezy} is a 2,048-core chip with 8 threads per core. The processor is designed for scientific HPC installations, but with high parallelism for dense linear algebra, it is also very capable for AI training and inference. 
\item The Tenstorrent Grayskull accelerator $\langle$Tenstorrent$\rangle$~\cite{gwennap2020tenstorrent} has a 10x12 array of Tensix cores, each of which is comprised of five RISC cores. These RISC cores can compute in int8, fp16, and bfloat16, and the floating point formats operate at a quarter of the performance of int8.  
\item Preferred Networks is another Japanese company, and it works with the Tokyo University on chip design projects. The PFN-MN-3 PCIe accelerator card $\langle$PFN-MN-3$\rangle$~\cite{gwennap2020tenstorrent} has four chips, each of which have four die and 32 GB of RAM. Each die has 2048 processing elements and 512 matrix arithmetic blocks, all of which compute at fp16 precision. Indications are that these cards are accelerators for one of the Japanese exascale supercomputers, but they also happen to have the arithmetic units for both training and inference. 
\end{itemize}

\subsubsection{FPGA-based Accelerators}

\begin{itemize}
\item The Intel Arria solution pairs an Intel Xeon CPU with an Altera Arria FPGA $\langle$Arria$\rangle$~\cite{hemsoth2018intel,abdelfattah2018dla}. The CPU is used to rapidly download FPGA hardware configurations to the Arria, and then farms out the operations to the Arria for processing certain key kernels. Since inference models do not change, this technique is well geared toward this CPU-FPGA processing paradigm. However, it would be more challenging to farm ML model training out to the FPGAs. Since it is an FPGA, the peak performance is equal to the performance the DNN model, the performance peak is reported for using GoogLeNet which ran at 900 fps. 
\item The Bittware/Achronix VectorPath S7t-VG6 accelerator $\langle$Achronix$\rangle$~\cite{roos2019fpga} is a FPGA-based processor on a PCI Express card. The FPGA includes eight banks of GDDR6 memory, a 2000GbE and a 4000GbE network interfaces, and 40,000 int8 multiply-accumulate units. 
\item Cornami has been developing a reconfigurable AI chip based on FPGA technology. Their FPGA-based prototype posted impressive performance in fp16 precision~\cite{cornami2020cornami}, but their ASIC has not yet taped out. 
\item The Flex Logix InferX X1 eFPGA/DSP accelerator card $\langle$FlexLogix$\rangle$~\cite{mehta2020performance} targets both signal processing and machine learning markets and supports int8, int16, Bfloat16 and fp16 precisions. It has 4,000 multiply-accumulate units and is programmable by TensorFlow Lite and ONNX.
\item The Microsoft Brainwave project $\langle$Brainwave$\rangle$~\cite{morgan2017drilling} is a programmable Intel Stratix 10 280 FPGA that was deployed as part of the Catapult project~\cite{chiou2017microsoft}. It is intended for re-programmable inference processing. 
\end{itemize}

\subsubsection{GPU-based Accelerators}

There are three NVIDIA cards and two AMD/ATI cards on the chart (listed respectively): the Volta architecture V100 $\langle$V100$\rangle$~\cite{volta2019nvidia,smith201816gb}, the Turing T4 $\langle$T4$\rangle$~\cite{kilgariff2018nvidia}, the Ampere architecture A100 $\langle$A100$\rangle$~\cite{krashinsky2020nvidia}, the MI8 $\langle$MI8$\rangle$~\cite{exxactcorp2017taking}, and MI60 $\langle$MI60$\rangle$~\cite{smith2018amd}. The V100, A100, MI8, and MI60 GPUs are pure computation cards intended for both inference and training, while the T4 Turing GPU is geared primarily to inference processing, though it can also be used for training.

\subsubsection{Dataflow Chips and Cards}

Dataflow processors are custom-designed processors for neural network inference and training. Since neural network training and inference computations can be entirely deterministically laid out, they are amenable to dataflow processing in which computations, memory accesses, and inter-ALU communications actions are ``placed-and-routed'' onto the computational hardware. 

\begin{itemize}
\item Alibaba Hanguang 800 accelerator $\langle$Alibaba$\rangle$~\cite{peng2019alibaba} posted the highest inference rate for a chip when it was announced in September 2019. Although ResNet-50 inference benchmarks were released, no peak performance capability was reported nor were any architecture details from which peak performance could be calculated so the peak performance number is their reported ResNet-50 result. 
\item Amazon Web Services has released a few details of its Inferentia chip $\langle$Inferentia$\rangle$~\cite{hamilton2018aws,cloud2020deep}. A peak performance has been published for int8 inference, and it can also compute at fp16 and bfloat16 for compatibility with common DNN models. As an accelerator, it is likely deployed on a card or daughterboard so power consumption is likely in the 150-300 W range.  
\item Baidu has announced an AI accelerator chip called Kunlun $\langle$Baidu$\rangle$~\cite{merritt2018baidu, duckett2018baidu}. There are two variants of the Kunlun: the 818-100 for inference and the 818-300 for training. These chips are aimed at low power data center training and inference and are deployed in Baidu's cloud service.  
\item The Cambricon dataflow chip $\langle$Cambricon$\rangle$~\cite{cutress2018cambricon} was designed by a team at the Institute of Computing Technology of the Chinese Academy of Sciences (IST-CAS) along with the Cambricon company, which came out of the university team. They published both int8 inference and float16 training numbers that are both significant, so both are on the chart. 
\item Google has released three versions of their Tensor Processing Unit (TPU)~\cite{jouppi2018domain}. The TPU1 $\langle$TPU1$\rangle$~\cite{teich2018tearing} is only for inference, but Google soon made improvements that enabled both training and inference on the TPU2 $\langle$TPU2$\rangle$~\cite{teich2018tearing} and TPU3 $\langle$TPU3$\rangle$~\cite{teich2018tearing}.   
\item GraphCore.ai has released their C2 card $\langle$GraphCoreC2$\rangle$~\cite{lacey2017preliminary,gwennap2020groq} in early 2019, which is being shipped in their GraphCore server node (see below). This company is a startup headquartered in Bristol, UK with an office in Palo Alto. They have strong venture backing from Dell, Samsung, and others. The performance values for inference benchmarking were achieved with ResNet-50 training for the single C2 card with a batch size for training of 8. 
\item The Groq Tensor Streaming Processor (TSP) $\langle$Groq$\rangle$~\cite{gwennap2020groq,abts2020think} is a single processor comprised of over 400,000 multiply-accumulate units along with memory and vector engines. They are organized into sets of Superlanes with each Superlane executing a very long instruction word (VLIW) instruction. Data shuttles between memory units and arithmetic units both within a Superlane and neighboring Superlanes to complete a static program of execution. Each processing unit operates on 8-bit words, and multiple adjacent processing units can be ganged together to execute floating point and multi-byte integer operations. The Groq TSP currently holds the record for most images per second for ResNet-50 processing. 
\item The Huawei HiSilicon Ascend 310 $\langle$Ascend310$\rangle$~\cite{huawei2020ascend310} is an integrated CPU with AI accelerator, which is aimed at autonomous systems. It is based on the same Huawei Da Vinci architecture as the data center focused Ascend 910. The Ascend 310 reports respectable performance for both int8 and fp16 and is intended for inference applications. The Ascend 910 $\langle$Ascend910$\rangle$~\cite{huawei2020ascend910} consumes over 300W of power and is the core of the Huawei AI Cluster product line. 
\item Intel purchased the Israeli startup Habana Labs in 2018, and Habana has released two chips. The Goya chip $\langle$Goya$\rangle$~\cite{gwennap2019habanagoya,medina2020habana} is an inference chip, and its peak performance is an estimate based on the size of the MAC array and 1.5 GHz clock speed typical of 7-nm processors. Habana Labs has also released a training processor called Gaudi $\langle$Gaudi$\rangle$~\cite{gwennap2019habanagaudi,medina2020habana}. Intel Habana is planning to release server appliances that include eight Goyas or eight Gaudis. 
\item The Cerebras CS-1 Wafer Scale Engine (WSE) $\langle$Cerebras$\rangle$~\cite{hock2019introducing,merrit2019startup} is the first wafer-scale processor; it has over 400,000 cores across 84 chips on the same wafer and draws a maximum power of 20kW. Each core has 45 Kbytes of local memory, and the cores are interconnected with a mesh network. While Cerebras has not released clock speeds, numerical precision, or computational performance numbers, there has been discussion that the WSE peak performance exceeds one petaflop. Hence, the graph shows an estimate based on an estimated clock speed of 2GHz for fp16 precision, and that all of the 400,000 cores can execute a fused-multiply-add (FMA) every clock cycle.  
\end{itemize}

\subsection{Data Center Systems}

This section lists a number of single-node data center systems. 

\begin{itemize}
\item There are four NVIDIA server systems on the graph: the DGX-Station, the DGX-1, the DGX-2, and the DGX-A100: The DGX-Station is a tower workstation $\langle$DGX-Station$\rangle$~\cite{alcorn2017nvidia} for use as a desktop system that includes four V100 GPUs. 
The DGX-1 $\langle$DGX-1$\rangle$~\cite{alcorn2017nvidia,cutress2018nvidias} is a server that includes eight V100 GPUs that occupies three rack units, while the DGX-2 $\langle$DGX-2$\rangle$~\cite{cutress2018nvidias} is a server that includes sixteen V100 GPUs that occupies ten rack units. Finally, the recently announced DGX-A100 $\langle$DGX-A100$\rangle$~\cite{campa2020defining} contains eight A100 GPUs that are networked together with a third generation NV-Link network. 

\item GraphCore.ai started shipping a Dell/EMC based DSS8440 IPU-Server $\langle$GraphCoreNode$\rangle$~\cite{graphcore2020dell} in 2019, which contains eight C2 cards (see above). The server power is an estimate based on the components of a typical Intel based, dual-socket server with 8 PCI cards. Training results are on ResNext-50~\cite{lacey2020updated}. 
\item The SolidRun Janux GS31 incorporates 128 Gyrfalcon Lightspeeur 5801 (see above) video processing chips $\langle$GyrfalconSaystem$\rangle$~\cite{hpcwire2020solidrun}, which uses a matrix processing engine with processor-in-memory techniques to compute model inference with a maximum power draw of 900W.  
\end{itemize}

\section{Announced Accelerators}

In this section, let us pivot towards the future. A number of other accelerator chips have been announced but have not published any performance and power numbers. Below are several companies from whom interesting announcements are expected in the next year or so:  
\begin{itemize}
\item Qualcomm has announced their Cloud AI 100 accelerator~\cite{mcgrath2019qualcomm}, and with their experience in developing communications and smartphone technologies, they have the potential for releasing a chip that delivers high performance with low power draws. 
\item SambaNova has been hinting at their reconfigurable AI accelerator technology from within stealth mode for a few years, but they have not provided any details from which we can estimate performance or power consumption of their solutions. 
\item Blaize has emerged from stealth mode and announced its Graph Streaming Processor (GSP)~\cite{yoshida2019blaize}, but they have not provided any details beyond a high level component diagram of their chip. 
\item Enflame has announced it's CloudBlazer T10 data center training accelerator~\cite{hpcwire2019enflame}, which will support a broad range of datatypes. 
\item Esperanto is building its Maxion CPU AI processor~\cite{gwennap2018esperanto} on the RISC-V open instruction set standard, but they have not released enough information on performance or power consumption to make an assessment. 
\end{itemize}

While there are a few neuromorphic and neuromorphic-inspired  chips in the above lists, there have also been a number of announcements from companies that intend to release such accelerators. Here is a list of them. 

\begin{itemize}
\item Brainchip Akida spiking neural network processor features 1024 neurons per chip, similar to the IBM TrueNorth research chip, that runs on less than one Watt of power~\cite{dr2020brainchip}. It is expected to use an 11-layer SNN. 
\item Eta Computing TENSAI Chip was demonstrated using spiking neural networks, but Eta Computing has since released a more conventional inference chip because they determined that spiking neural network chips were not ready for commercial release~\cite{moore2020low}. It is not clear whether the TENSAI chip will be released commercially. 
\item aiCTX (pronounced AI cortex) is developing a low-power, low-latency neuromorphic accelerator~\cite{eetimes2018baidu} that executes single sample inference in under a milliwatt and under 10 ms. 
\item Anaflash chip is an eflash-based spiking neuromorphic chip~\cite{clarke2018ai} that encodes 320 neurons with 68 interconnecting synapses per layer. The Univ. of Minnesota research paper that explains this technology is~\cite{kim2019parallel}, and it provides results on the MNIST written handwriting dataset. 
\item The Grai Matter Labs chip called NeuronFlow has 1024 neurons per chip that can operate with 8-bit or 16-bit integer precision~\cite{dahad2019startup,clark2020grai}.
\item The startup Koniku is modeling circuitry after the brain by designing a co-processor built with biological neurons~\cite{agabi2016cell}. The core of the device is a structured micro electrode array system (SMEAS) that they call a Konikore. They have demonstrated that keeping neurons alive is a solvable engineering control problem: living neurons operate in a defined parameter space, and they are developing hardware and algorithms which control the the parameter space of the environment.
\item Finally, the Intel Loihi chip~\cite{hemsoth2018first} is a spiking neural network processor that has been scaled up to 768 chips to simulate 100 million spiking neurons. 
\end{itemize}

Several companies have made technology announcements to use optical processing and silicon photonic networks for AI processing~\cite{dunietz2017light}. These companies include Intel~\cite{feldman2019silicon}, Light Matter~\cite{feldman2018photonic}, Lighton~\cite{lighton2020photonic}, Lightelligence~\cite{shen2017deep,clarke2019startup}, Optalysys~\cite{feldman2018optalysys}, Fathom Computing~\cite{feldman2018optical}, and Luminous~\cite{giles2019bill}. 

As performance and power numbers become available for all of these and other chips, they will be added in future iterations of this survey. 

And, as one would expect in such a tumultuous innovation environment, there have also been companies that have cancelled their AI accelerator programs and even a few that have already declared bankruptcy. The most prominent of these was Intel's halt in development of the Nervana NNP-T training chips (codenamed Spring Crest) and curtailing development of the Nervana NNP-I inference (codenamed Spring Hill) in January 2020~\cite{freund2020intel}. 
A little further in the past, KnuEdge, the company that announced the KnuPath chip~\cite{clarke2016military} shut its doors in 2018, while Wave Computing, which also owns the MIPS processor technology, folded in early 2020. Finally, TeraDeep, a startup from some Purdue University professors and researchers closed its doors somewhere around 2016 before much of the AI chip race started.

\section{Summary}

This paper updated the survey of deep neural network accelerators that span from extremely low power through embedded and autonomous applications to data center class accelerators for inference and training. Many more accelerators were announced and released, and many of these releases included peak performance and power consumption data. There has been a drive to release 8-bit integer performance for edge/embedded accelerators, while data center accelerators for training often presented 16-bit float performance data. Finally, the diversity of architecture and technologies including neuromorphic, flash-based analog memory processing, dataflow engines, and photonic-based processing is making the competition and performance opportunities very exciting.

\section*{Acknowledgement}

We are thankful to Masahiro Arakawa, Bill Arcand, Bill Bergeron, David Bestor, Bob Bond, Chansup Byun, Vitaliy Gleyzer, Jeff Gottschalk, Michael Houle, Matthew Hubbell, Anna Klein, David Martinez, Lauren Milechin, Sanjeev Mohindra, Paul Monticciolo, Julie Mullen, Andrew Prout, Stephan Rejto, Antonio Rosa, Charles Yee, and Marc Zissman
for their support of this work. 


\bibliographystyle{IEEEtran} 
\bibliography{MLAccelerators}


\end{document}